\documentclass[vecphys]{svmult}

\usepackage{makeidx}         
\usepackage{graphicx}        
\usepackage{multicol}        
\usepackage[bottom]{footmisc}

\makeindex             

\begin{document}

\title*{Star Formation in the Central Regions of Galaxies}
\author{J.~H. Knapen\inst{1}\and
E.~L. Allard\inst{1}\and
L.~M. Mazzuca\inst{2}\and
M. Sarzi\inst{1}\and
R.~F. Peletier\inst{3}}
\institute{University of Hertfordshire, Centre for Astrophysics
  Research, Hatfield, Herts AL10 9AB, UK,
\texttt{j.knapen@herts.ac.uk}
\and NASA Goddard Space Flight Center, Greenbelt, MD 20771, USA
\and University of Groningen, NL-9700 AV Groningen, The Netherlands}
%
%
\maketitle

\begin{abstract}

Massive star formation in the central regions of spiral galaxies plays
an important role in the dynamical and secular evolution of their
hosts.  Here, we summarise a number of recent investigations of the
star formation history and the physical conditions of the gas in
circumnuclear regions, to illustrate not only the detailed results one
can achieve, but also the potential of using state-of-the-art
spectroscopic and analysis techniques in researching the central
regions of galaxies in general.  We review how the star formation
history of nuclear rings confirms that they are long-lived and stable
configurations. Gas flows in from the disk, through the bar, and into
the ring, where successive episodes of massive star formation
occur. Analysing the ring in NGC~7742 in particular, we determine the
physical conditions of the line emitting gas using a combination of
ionisation and stellar population modelling, concluding that the
origin of the nuclear ring in this non-barred galaxy lies in a recent
minor merger with a small gas-rich galaxy.

\end{abstract}

\section{Introduction}
\label{intro}

Starbursts, defined as relatively short periods of enhanced massive
star formation (SF) activity, are important events in the evolution of
galaxies. They enhance the luminosity of the host galaxy, facilitating
the detection of those at larger distances, can transform significant
quantities of gas into stars, can cause metal injection into, and
mixing of, the interstellar medium, and can help the secular evolution
of galaxies (e.g., reviews by Kormendy \& Kennicutt 2004; Gallagher
2005). We focus here on a particular class of low-luminosity
starbursts occurring in nuclear rings in spiral galaxies, which allow
the detailed study of their SF histories and of the physical
properties of the gas in the circumnuclear region.

\section{Cold gas flowing in and feeding the starburst}
\label{coldgas}

The centres of many spiral galaxies show evidence of massive SF, often
organised into a nuclear ring or pseudo-ring. Such rings are found in
20{$\% $} of spiral galaxies (Knapen 2005), are thus relatively
common, and almost always occur within a barred host (Knapen
2005). Dynamically, nuclear rings are thought to trace the position of
the inner Lindblad resonances (ILRs), where gas driven in under the
influence of the large bar slows down (Knapen et al. 1995, hereafter
K95, and references therein).  M100 (NGC~4321) is a prominent,
relatively face-on spiral galaxy with a moderately strong bar, at a
distance of 16.1~Mpc (so 1\,arcsec corresponds to 70\,pc). M100 hosts
a well-known nuclear ring with prominent massive SF, which is located
near a pair of ILRs induced by the bar (K95).

\begin{figure}
\centering
\includegraphics[height=4.5cm]{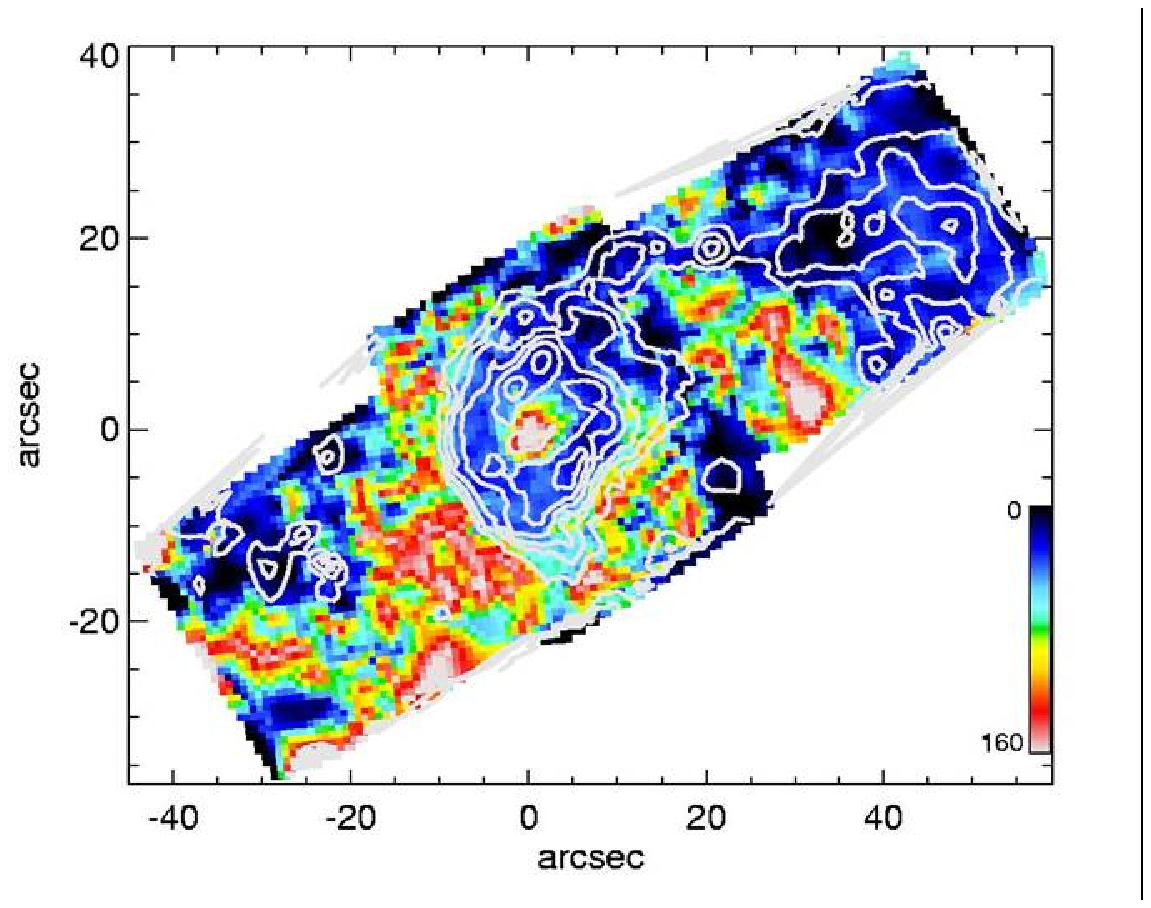}\includegraphics[height=4.5cm]{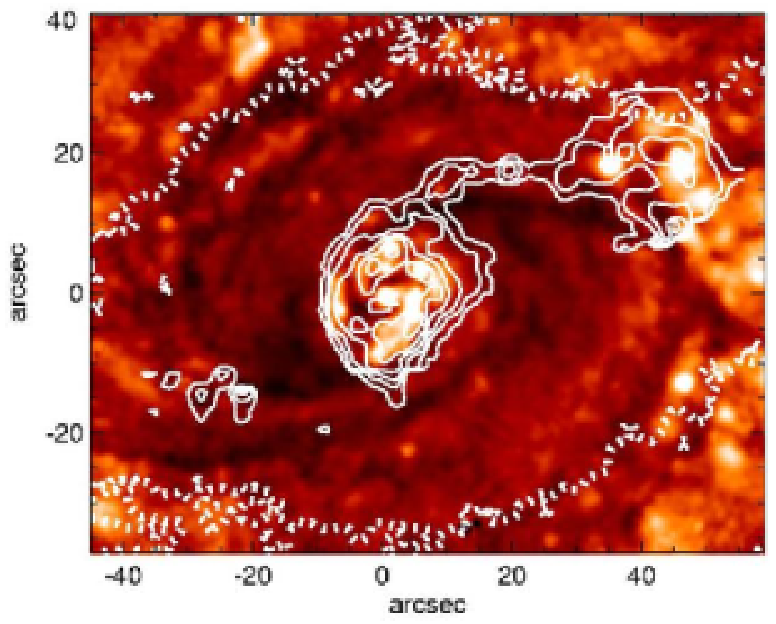}
%
%
\caption{{\it Left:} SAURON gas velocity dispersion in the central
  region of M100 (in km\,s$^{-1}$), with, overlaid, contours of
  H{$\beta$} line emission at relative levels [0, 0.05, 0.1, 0.2, 0.5,
  1, 2, 3]. {\it Right:} \emph {B-R} image with the location of the
  bar indicated by a $K_{\rm s}$ band contour (thick dashed line) at
  18.3\,mag\,arcsec$^{-2}$; H{$\beta$} emission line contours as in
  the {\it left} panel. Reproduced with permission from Allard et
  al. (2005).}
\label{coldgasfig}       
\end{figure}

We have obtained integral-field spectroscopic data with SAURON (Bacon
et al. 2001) on the William Herschel Telescope (WHT) of the central
region of the barred spiral galaxy M100 (Allard et al. 2005, 2006).
These data allow us to derive maps of emission line intensities, line
strength indices, and the gas velocity dispersion across the
circumnuclear region. Fig.~\ref{coldgasfig} shows how the H{$\beta$}
emission traces the nuclear ring, broken into hotspots of emission.

The gas velocity dispersion map (Fig.~\ref{coldgasfig}, {\it left}
panel) shows clearly that the ring has lower dispersion than the
underlying disk, and is thus cooler, exactly where the strongest
H{$\beta$} emission occurs. This happens within the ring, but in the
bar the cold gas is aligned and offset from the dustlanes
(Fig.~\ref{coldgasfig}, {\it right} panel). This is confirmation of
the gas inflow model of nuclear ring formation, as the low dispersion
traces cold gas flowing into the area through the dustlanes, and
accumulating into the ring under the influence of the
ILRs. Instabilities within this gas then trigger significant massive
SF (Allard et al. 2005).

\section{Star formation histories: evidence for multiple bursts}
\label{sfhist}

\subsection{M100}
\label{m100}

The SAURON observations described in the previous Section not only
allow analyses of the kinematics and of the emission line
characteristics, but also an in-depth study of the SF history of the
circumnuclear region of M100 (Allard et al. 2006). For this, we
extracted H$\beta$, Mg{\emph b} and Fe5015 absorption line indices
across the field of view, where the latter two were subsequently
combined into one, MgFe$= \sqrt{\mbox{Mg{\emph
b}}\cdot\mbox{Fe5015}}$, which minimises the effect of differing
abundance ratios across the galaxy (Falc\'on-Barroso et al. 2002). The
indices vary with time for a single burst stellar population (SSP):
the H$\beta$ absorption line index first rises, reaching a peak at an
age of around 250\,Myr, then decreases gradually, whereas the MgFe
index rises monotonically with age.

\begin{figure}[h]
\centering
\includegraphics[height=9.5cm]{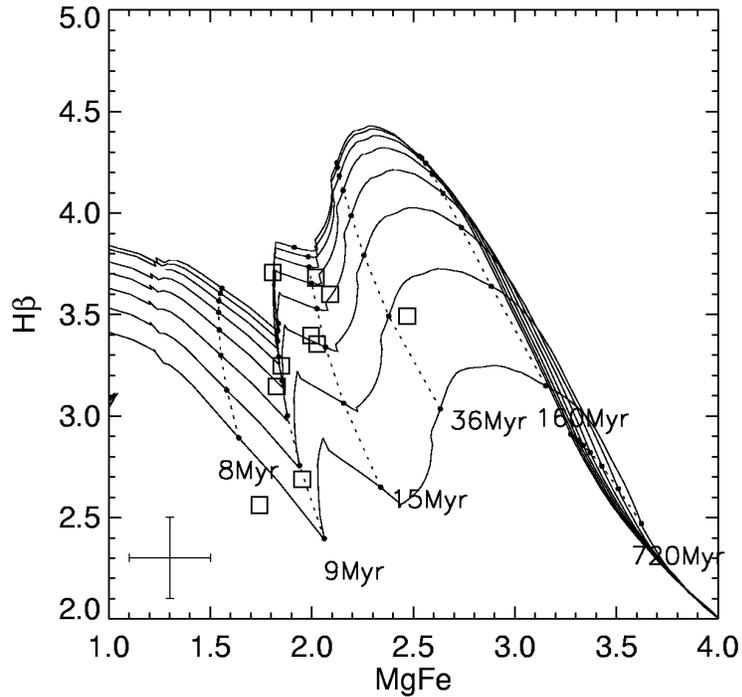}
%
%
\caption{Measured H$\beta$ and MgFe indices for different points in
  the nuclear ring of M100 (squares), compared to a family of models
  which give the time evolution (indicated, in Myr) of the indices in
  an SSP. The models are combinations of an old
  ($\sim3$\,Gyr) bulge/disk population with one (bottom) to eight (top)
  discrete SF episodes within the last $\sim$\,0.5\,Gyr, separated by
  100\,Myr. From Allard et al. (2006).}
\label{m100model}       
\end{figure}

Comparing SSP model predictions with the data points measured for the
nuclear ring and centre of M100 shows that whereas the nucleus can be
fitted well and yields an age of some 3\,Gyr, the nuclear ring points
cannot be fitted at all because the H$\beta$ index values are much
lower then expected for any reasonable SSP (Allard et al. 2006). We
thus introduce a family of composite models, which are characterised
by a combination of an old underlying bulge and/or disk contribution
and a series of much more recent SF events. We estimate the age of the
bulge/disk as 3\,Gyr from our fit of an SSP model to the nuclear data
point. Adding a number of recent bursts, 100\,Myr apart, and starting
a couple of hundred Myr ago so that the latest burst occurs now (to
yield the observed Balmer line emission) then produces a grid of model
lines, illustrated in Fig.~\ref{m100model}, which straddle the
observed nuclear ring data points.

Our proposed model (Allard et al. 2006) not only fits the data, but is
also phenomenologically attractive: it stipulates the formation of the
underlying bulge/disk component and the current SF activity in the
ring, but also the long-lived nature of the ring, albeit not
constantly forming stars at the same high rate as observed now. Our
modelling thus confirms a picture in which the bar in M100, a stable
configuration, constantly channels gas into the nuclear region, where
it concentrates in the nuclear ring until it reaches a density high
enough to collapse gravitationally, and enter the next starburst
phase. That phase will last a relatively short time, until either the
gas runs out or the negative feedback from the massive SF stops the
formation of further stars, after which the process of accretion
followed by instability to SF can start again.

\subsection{Other nuclear rings}
\label{more_rings}

The model described above for M100 yields a good fit to the data, as
well as a convincing interpretational framework which ties the origin
and dynamics of the nuclear ring to its SF properties. M100 is often
seen as a prototype of a barred galaxy hosting a nuclear ring, and to
confirm this further, we analysed the SF history of a sample of seven
more nuclear rings (Allard et al. 2007). This analysis is based on
long-slit spectra obtained with the ISIS spectrograph on the WHT,
covering a wide wavelength range from which we could extract values
for the H$\beta$, MgFe, H$\delta$A, and D(4000) line indices for
selected points in the rings, as well as in the nuclei of the host
galaxies.

We find that the nuclear data points can be reproduced well by a
single old SSP model, yielding, as in the case of M100, ages of a
couple of Gyr for the underlying bulge and/or disk component. In none
of the eight sample galaxies the nuclear ring data points can be
fitted by such a model, reproducing again the results for M100
(Sect.~3.1; Allard et al. 2006).

\begin{figure}[h]
\centering
\includegraphics[height=6cm]{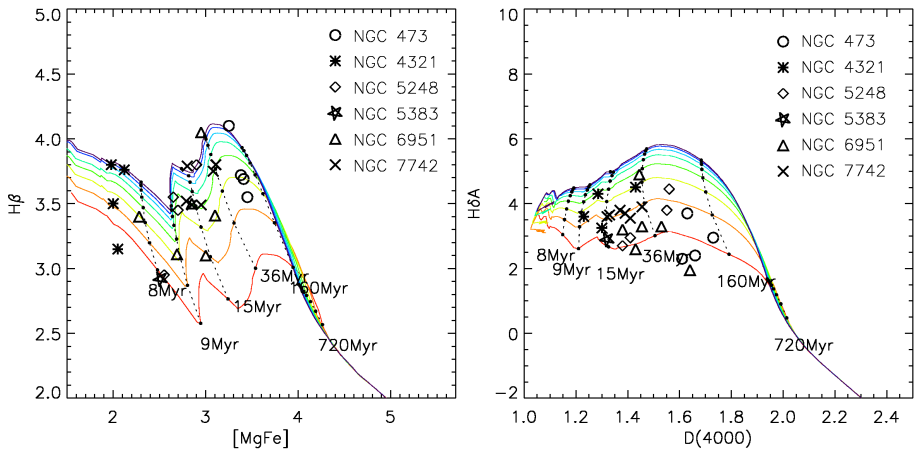}
%
%
\caption{Evolution of the indices H$\beta$ vs. MgFe ({\it left}) and
  H$\delta$A vs. D(4000) ({\it right}) for a family of eight composite
  stellar population models, compared to the nuclear ring data points
  measured from ISIS long-slit spectra for six galaxies. The different
  model lines refer to models with one (bottom) to eight (top)
  discrete SF episodes within the last $\sim$\,0.5\,Gyr, separated by
  100\,Myr. From Allard et al. (2007).}
\label{isismodel}       
\end{figure}

Figure~\ref{isismodel} shows how we can fit the data points for six of
the galaxies with composite SSP models very similar to the ones
developed for M100. We can thus generalise our conclusion that most
nuclear rings are stable configurations, with recurrent episodes of
massive SF, the latest of which we are witnessing. The use of the
H$\delta$A vs. D(4000) plot (Fig.~\ref{isismodel}), made possible by
the wider wavelength range of the ISIS observations compared to the
SAURON data set, allows us to exclude with a high degree of certainty
a model in which the nuclear ring is only a one-off massive SF event,
preceded merely by the SF which formed the bulge and/or disk, some
3\,Gyr ago. Such a model could, at a stretch, still explain the M100
SAURON data, but can now be excluded.

The galaxies NGC~4314 and NGC~7217 can be explained by a model
representing an old underlying bulge/disk component, of some 10\,Gyr
old, and only one, current, burst of massive SF. This would imply that
the nuclear rings in these galaxies are new to forming massive stars,
at least at their current position. So either they have only now
entered the first episode where their gas clouds are unstable against
gravitational collapse, or the rings have shrunk, and previous SF
events have happened at slightly larger radii (Benedict et al. 2002),
not covered by our data points as plotted in Fig.~\ref{isismodel}
(Allard et al. 2007).

\section{Physical conditions of the gas: the case of NGC~7742}
\label{physcond}

Another important aspect in deciphering the history and properties of
nuclear rings, and, perhaps more importantly, in recognising the role
they can play in tracing the evolution of their host galaxy, is the
study of the physical properties of the ionised gas. This can be done
by using diagnostic emission line diagrams, such as those introduced
by Veilleux \& Osterbrock (1987), which plot the ratio between two
blue emission lines against that between two red lines, ensuring that
the effects of dust on the diagnostics are limited.

We have recently started a programme to use integral-field
spectroscopy (IFS) measurements of the circumnuclear regions of
galaxies to disentangle not just their SF histories, but also the
physical conditions of the gas, in order to constrain the history of
these regions and to model the interplay between them and their host
galaxies. As a first step, and as a feasibility study, we have
combined blue (SAURON, from Falc{\'o}n-Barroso et al. 2006) and red
(DensePak on the WYIN telescope) IFS data of the nuclear ring region
in the non-barred galaxy NGC~7742 (Mazzuca et al. 2006). The fields of
view of the instrument cover the ring nicely, and the combination of
wavelength ranges means that we can deduce information on most of the
lines traditionally used in emission line diagnostics.

NGC~7742 is one of the few known cases where the galaxy hosting a
star-forming nuclear ring contains no significant bar. The origin of
such rings may lie in a previously existing, but now dissolved, bar,
in a weak oval, or in a past or present interaction or minor merger
(see Knapen et al. 2006 for a literature overview and
examples). NGC~7742 is known to have counterrotating gas and stars in
the central region (e.g., de Zeeuw et al. 2002), which immediately
points to an interactive past.

\begin{figure}[h]
\centering
\includegraphics[height=6cm]{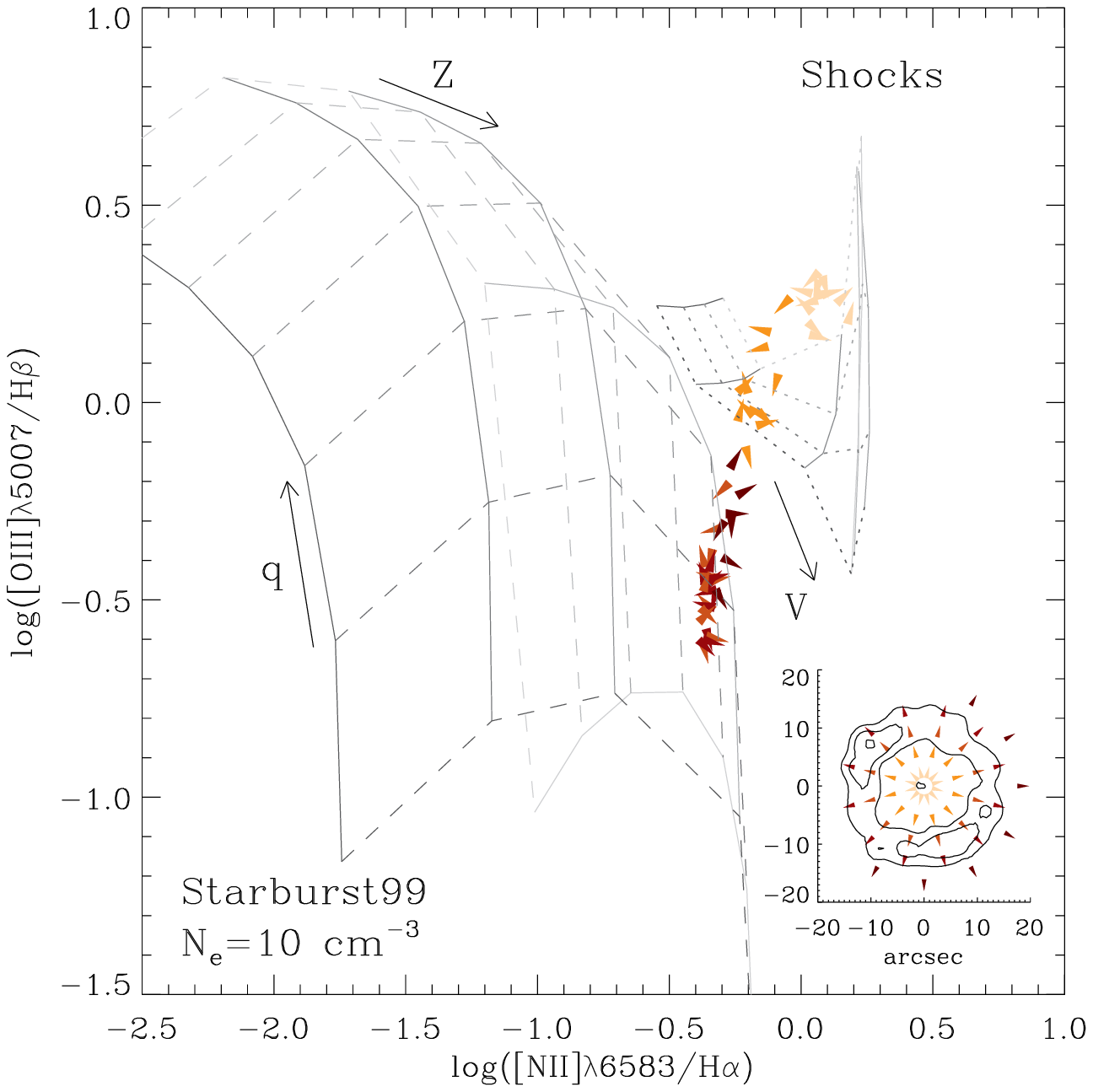}\includegraphics[height=6cm]{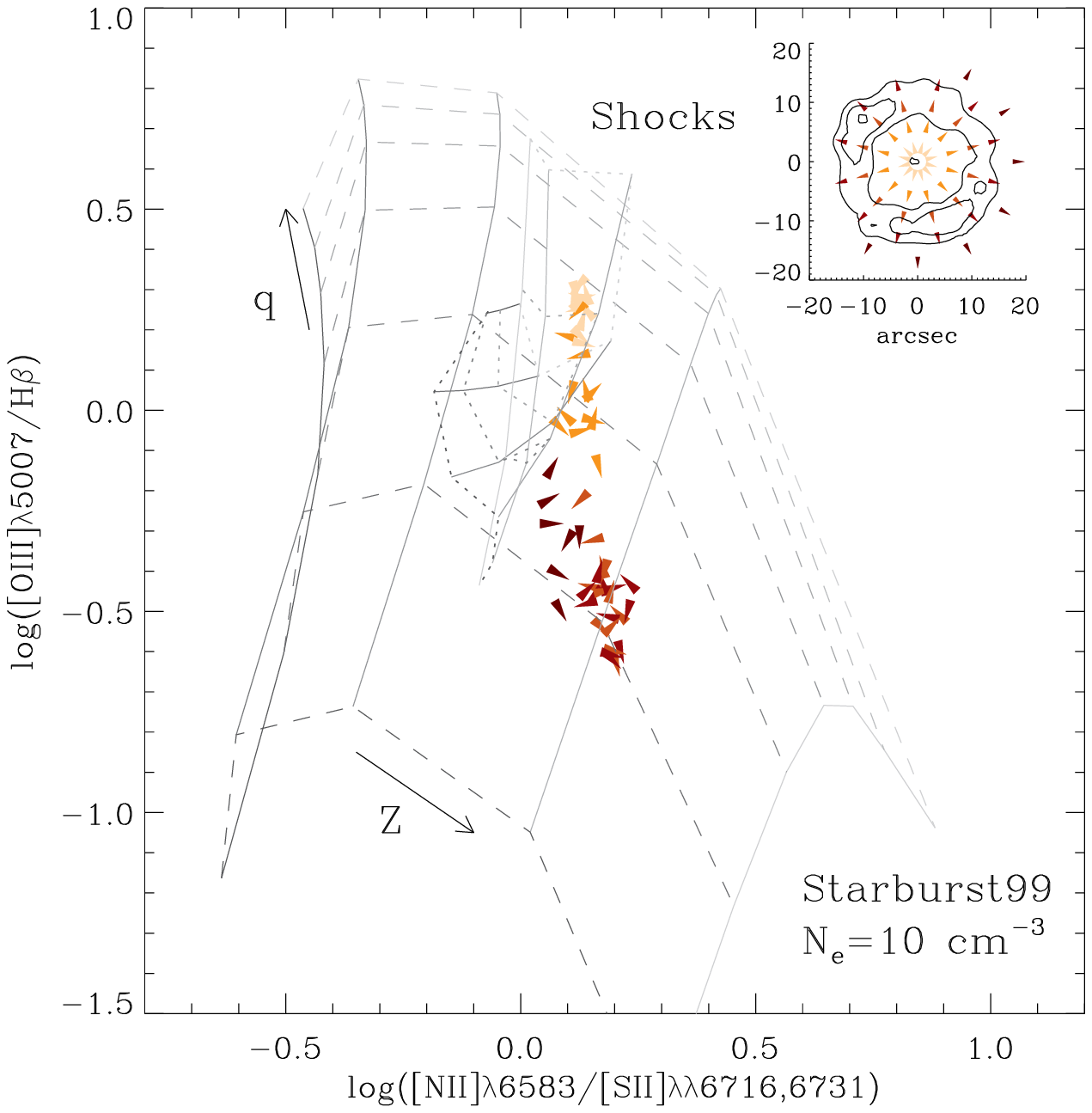}
%
%
\caption{Diagnostic diagrams of [O{\sc iii}]$\lambda5007$/H$\beta$
 vs. [N{\sc ii}]$\lambda6583$/H$\alpha$ ({\it left}) and [O{\sc
 iii}]$\lambda5007$/H$\beta$ vs. [N{\sc ii}]$\lambda6583$/[S{\sc
 ii}]$\lambda\lambda6716,6731$ ({\it right}). Data points are shown by
 filled triangles, and the inset shows the location of these points in
 the ring region, with the contours showing the H$\alpha$ emission of
 the circumnuclear ring.  Darker symbols refer to points at larger
 radii; they are oriented according to their position angles. In each
 panel, MAPPINGS III starburst or shock models are shown by the grid
 of solid and dashed lines or of solid and dotted lines,
 respectively. Reproduced with permission from Mazzuca et al. (2006).}
\label{lineratios}       
\end{figure}

Our IFS data yield, first of all, a low value for the electron
density, $N_{\rm e}$, of order 10\,cm$^{-3}$, in the ring region, as
derived from the ratio of the [S{\sc ii}] lines at 6716 and 6731\,\AA\
(assuming $T=10^4$\,K, see Mazzuca et al. 2006). The knowledge of this
value allows us to model other line ratios, such as those of [O{\sc
iii}]\,$\lambda$5007\,\AA/H$\beta$, [N{\sc
ii}]\,$\lambda$6583/H$\alpha$, or [N{\sc ii}]\,$\lambda$6583/[S{\sc
ii}]\,$\lambda\lambda$6716,\,6731\AA, as shown in
Fig.~\ref{lineratios}. In that figure, the different points in and
around the ring are indicated by shade and shape, as reproduced in the
inset which shows an H$\alpha$ image of the region, and with the
points with the lowest [O{\sc iii}]\,$\lambda$5007\,\AA/H$\beta$
ratios corresponding to those located in the ring.

The left panel of Fig.~\ref{lineratios} shows a comparison with a set
of MAPPINGS\,III shock and SF models (Kewley et al. 2001), which
confirms that SF is the dominant ionisation mechanism within the ring,
but that shock ionisation is important both in- and outside the
ring. It also indicates that there is little azimuthal variation in
line ratios around the ring. The right panel of Fig.~\ref{lineratios}
yields valuable information on the metallicity, which we find to be
very near solar in the nuclear ring region (as indicated by the drawn
line going through the nuclear ring points, which is at $Z=Z_\odot$).

So what conclusions can we draw about the origin of the nuclear ring?
The non-barred nature of its host, NGC~7742, and the counterrotation
observed in the central region might indicate that some past
interactive event is responsible. For the similar non-barred nuclear
ring host galaxy NGC~278, a past minor merger is indeed what we
proposed as origin for its ring on the basis of the severely disturbed
morphology and kinematics of its atomic gas (Knapen et
al. 2004). NGC~7742 has a companion galaxy, NGC~7743, but that seems
too far away and of too regular morphology to be a candidate for a
recent interaction. A minor merger with a small gas-rich galaxy, as in
the case of NGC~278, seems the most likely, but this would undoubtedly
lead to an influx of low-$Z$ gas, seemingly at odds with our
near-solar metallicity measurement.

Here is where the SF history models as described above (Sect.~3) come
to our aid: the multiple bursts of SF, for which we found evidence
there, would have enriched the gas to near-solar metallicity on
timescales which are completely consistent with those modelled above
(Mazzuca et al. 2006). We thus conclude that indeed a past minor
merger has caused the asymmetry in the gravitational potential which
lies at the origin of the nuclear ring. To confirm this further, we
are presently pursuing H{\sc i} observations of NGC~7742: our model
predicts non-regular H{\sc i} morphologies and velocity fields.

\section{Closing remarks}
\label{closing}

The results highlighted here illustrate the use of integral-field
spectroscopic analysis techniques to study key aspects of
circumnuclear regions of galaxies: their detailed SF histories, and
their physical conditions. We have found strong evidence for the
stability and longevity of nuclear rings, which we model to form
massive stars episodically, for the past 0.5\,Gyr or so. Using the
nuclear ring in the non-barred galaxy NGC~7742 as a showcase for our
techniques, we find low electron densities, SF-dominated ionisation
and near-solar metallicity in the ring. We postulate that a minor
merger with a small gas-rich galaxy has led to the nuclear ring, and
that the episodic SF in the ring, as modelled above, has enriched the
infallen metal-poor gas to its current metallicity.

Our results show the importance of nuclear rings, being stable,
long-lived structures which transform significant amounts of disk gas
into inner-kpc stars. The techniques we have outlined here can be
applied more widely than to nuclear rings, though, and should lead to
important insights on, e.g., nuclear starbursts, or the zones
surrounding AGN or composite nuclei.

\hspace{2mm}

{\it Acknowledgements} We thank our co-workers on the papers as
reviewed here. JHK thanks the Leverhulme Foundation for the award of a
Leverhulme Research Fellowship, and the Royal Society for the award of
a conference grant which allowed him to attend the meeting in
Ishigaki.



\printindex
\end{document}